\documentclass[aps,pre,twocolumn, titlepage]{revtex4-2}
\usepackage{graphicx}
\usepackage{color}
\usepackage{dcolumn}
\usepackage{amsmath}
\usepackage{amsfonts}
\usepackage{bm}
\usepackage{epstopdf}
\usepackage{tabularx} 
\usepackage{multirow, booktabs}
\begin{document}

\title{Shapiro steps observed in a two-dimensional Yukawa solid modulated by a one-dimensional vibrational periodic substrate }

\author{Zhaoye Wang$^{1}$}
\author{Nichen Yu$^{1}$}
\author{C.~Reichhardt$^{2}$}
\author{C.~J.~O.~Reichhardt$^{2}$}
\author{Ao Xu$^{1}$}
\author{Xin Chen$^{1}$}
\author{Yan Feng$^{1,}$}
\thanks{The author to whom correspondence may be addressed: fengyan@suda.edu.cn}
\affiliation{$^{1}$ Institute of Plasma Physics and Technology, Jiangsu Key Laboratory of Frontier Material Physics and Devices, School of Physical Science and Technology, Soochow University, Suzhou 215006, China\\
{$^{2}$ Theoretical Division, Los Alamos National Laboratory, Los Alamos, New Mexico 87545, USA}
    }

\date{\today}

\begin{abstract}
Depinning dynamics of a two-dimensional (2D) solid dusty plasma modulated by a one-dimensional (1D) vibrational periodic substrate are investigated using Langevin dynamical simulations. As the uniform driving force increases gradually, from the overall drift velocity varying with the driving force, four significant Shapiro steps are discovered. The data analysis indicate that, when the ratio of the frequency from the drift motion over potential wells to the external frequency from the modulation substrate is close to integers, dynamic mode locking occurs, corresponding to the discovered Shapiro steps. Around both termini of the first and fourth Shapiro steps, the transitions are found to be always continuous, however, the transition between the second and third Shapiro steps is discontinuous, probably due to the different arrangements of particles.

\end{abstract}

\maketitle

\section{Introduction}

Collective dynamics of interacting particles under substrate modulation have been extensively studied in various physical systems, such as colloids~\cite{CReichhardtPRE:2005}, vortex lattices in type-II superconductors~\cite{KHaradaScience:1996}, pattern-forming systems~\cite{ASenguptaPRB:2010}, Wigner crystals~\cite{MCChaPRL:1998}, and dusty plasmas~\cite{LIWPRE:2019}. While driven by a uniform force with different magnitudes, these substrate-modulated systems exhibit interesting depinning dynamics~\cite{LIWPRE:2019, CReichardtRPP:2017}, e.g., three dynamical states of the pinned, disordered plastic flow, and moving ordered can be observed. Besides these three dynamical states, depinning dynamics also include more interesting phenomena, including kinks and antikinks~\cite{TBohleinNM:2012}, superlubricity or the Aubry transition~\cite{DMandelliPRB:2012}, directional locking~\cite{CReichardtPRE:2004}, and Shapiro steps~\cite{MPNJunipernc:2015}. 

Shapiro steps are one of prominent characteristics commonly observed in nonlinear depinning dynamical systems with competing timescales~\cite{youfengweiPRE:2022}. When particles move over a periodic substrate by a uniform driving force, the drift velocity of the particles increases with the driving force, so does the frequency of the drift motion over potential wells. When an additional alternating current (AC) frequency is introduced into the driving force or the substrate, the coupling between these two frequencies leads to new dynamics phenomena~\cite{MPNJunipernc:2015}. Especially, when the introduced AC frequency is synchronized with the frequency of the drift motion over potential wells, particles' drift velocity remains constant within certain driving force intervals due to resonance~\cite{TekicPRE:2010}, forming a series of steps known as the Devil's staircases~\cite{MHJensenPRL:1983} or the Shapiro steps~\cite{MPNJunipernc:2015}. The Shapiro steps have been observed in various systems, such as Josephson junctions~\cite{YuMShukrionvpRB:2013, VPSpRL:2024}, vortices in type-II superconductors~\cite{PMartinoliPRL:1990, LVanLookPRB:1999, CReichardtPRB:2000}, charge-density waves~\cite{GGrunerRMP:1988, REThornePRB:1988}, colloids~\cite{MPNJunipernc:2015}, skyrmions~\cite{CReichardtPRB:2015}, and Frenkel-Kontorova models~\cite{BHuPRE:2007}.

As a prominent experimental model system, dusty plasma, also termed as complex plasma, is composed of ions, electrons, neutral gas atoms, and micron-sized dust particles~\cite{JBeckersPOP:2023, H.M.ThomasNature:1996, BonitzPRL:2006, NosenkoPRL:2004, HKahlertPRL:2012, CDuPRL:2019, LIscience:1996, AMelzerPRE:1996, JHChuPRL:1994, HThomasPRL:1994, RLMerlinoPhysT:2004, VEFortovPhysR:2005, GEMorfillRevModPhys:2009, ApielPlasmaPhysics:2010, MbonitzRRP:2010, LiangcPRR:2023, YFengPRL1:2008, EThomasPhysPlasmas:2004, WYUPRE:2022, FWiebenPRL:2019, YHePRL:2020}. Under the typical laboratory conditions, these micron-sized dust particles are highly charged to $\approx 10^{4}$ elementary
charges negatively in plasma, interacting with each other through the Yukawa repulsion~\cite{KonopkaPRL:2000} of $\phi_{ij}$ = $Q^2{\rm exp}(-r_{ij}/\lambda_{\rm D})/4\pi \epsilon_{\rm 0} r _{ij}$, where $Q$ is the charge on each particle, $r_{ij}$ is the distance between two particles, and $\lambda _{\rm D}$ is the Debye screening length. These charged dust particles can be suspended and confined by the electric field in the plasma sheath, self-organizing into a single-layer, i.e., a two-dimensional (2D) dusty plasma~\cite{YFengPRE:2011, KQiaoPRE:2014}. Due to their high charges, these dust particles are strongly coupled to each other, exhibiting the typical solid~\cite{YFengPRL1:2008, PHartmannPRL:2014, AMelzerPRL:2012} and liquid like~\cite{PHartmannPRL:2014, YFengPRL2:2010, AMelzerPRL:2012} properties. Using various manipulations, such as powerful laser beams~\cite{YFengPRL1:2008, YFengPRL3:2010} or stripe electrodes~\cite{KJiangPOP:2009}, the dynamics of these dust particles can be easily modulated in experiments. While moving in plasma, these dust particles also experience a weak frictional gas damping~\cite{BLiuPRL:2003}. In 2D dusty plasma experiments, the motion of dust particles can be recorded by video imaging~\cite{YFengPRL1:2008, YFengPRL3:2010} and then accurately analyzed using particle tracking velocimetry~\cite{YFengRSI:2007, YFengRSI:2011, YzengRSI:2022, YFeng:2016}. Thus, dusty plasma can be used to study various fundamental physical procedures of solids and liquids at the individual particle level, including shocks~\cite{AKananovichPRE:2020}, diffusion~\cite{BLIUPRL:2008}, phase transitions~\cite{AMelzerPRE:1996, FengPRE:2008, DHuangPRR:2023, NYuPRE:2024}, and internal friction~\cite{sLuPRR:2023}.

Recently, the depinning dynamics of 2D dusty plasmas modulated by periodic substrates have been intensively investigated using computer simulations~\cite{LIWPRE:2019, zhuwenqiPRE:2022, L.GuPRE:2020, huangyuPRE2:2022, huangyuPRE1:2022}. From Ref.~\cite{LIWPRE:2019}, when a gradually increasing uniform driving force is applied to substrate-modulated 2D dusty plasmas, the pinned state, the disordered plastic flow, and the moving ordered state are all observed. Besides these three dynamical states, other dynamical behaviors are also found, including the directional locking~\cite{zhuwenqiPRE:2022}, the superlubric-pinned transition~\cite{huangyuPRE2:2022}, and the bidirectional flow~\cite{LIWPRR:2023}. However, from our literature search, the Shapiro steps have not been found in the previous investigations of dusty plasmas or Yukawa systems, as we study here.

The rest of this paper is organized as follows. In Sec.~\ref{sec2}, we briefly describe our computer simulation method of 2D solid dusty plasma modulated by a 1D vibrational periodic substrate and also driven by a uniform force. In Sec.~\ref{sec3}, we present our discovered Shapiro steps in the substrate-modulated 2D Yukawa solid. We also investigate the continuous/discontinuous transitions around both termini of each Shapiro step, using various structural and dynamical diagnostics. Finally, we provide a summary of these findings in Sec.~\ref{sec4}.

\section{Simulation methods}\label{sec2}

To characterize our studied 2D dusty plasmas, we follow~\cite{GJKalmanPRL:2004} to use two dimensionless parameters, which are the coupling parameter $\Gamma$ and the screening parameter $\kappa$~\cite{GJKalmanPRL:2004, H.OhtaPP:2000, K.Y.Sanbonmatsu:2001, GEMorfillRevModPhys:2009}. They are defined as $\Gamma = Q^2 / 4 \pi \epsilon_{0} a k_{\rm B} T $~\cite{GJKalmanPRL:2004} and $\kappa = a / \lambda_{\rm D}$~\cite{GJKalmanPRL:2004}, respectively. Here, $T$ is the kinetic temperature of particles, $a = (n \pi)^{-1/2} $ is the Wigner-Seitz radius for the areal number density $n$. Clearly, $\Gamma$ can be regarded as the inverse of the kinetic temperature. Note that, besides $a$, the lattice constant $b$ is also used as the length unit, which is $\approx 1.90 a$ for our studied 2D triangular lattice~\cite{GJKalmanPRL:2004}. 

To investigate the Shapiro step in substrate-modulated 2D dusty plasmas, we perform Langevin dynamical simulations~\cite{LIWPRE:2019}. In our simulated 2D dusty plasmas with $N_{\rm p}$ particles, the equation of motion for each particle $i$ is
\begin{equation}\label{EQyundong}
	m\mathbf{\ddot{r}}_i =- \sum \nabla \phi_{i j} - \nu m \mathbf{\dot{r}}_i + \mathbf{\zeta}_{i}(t) +\mathbf{F_{\rm s}} + \mathbf{F_{\rm d}}.
\end{equation}
Here, the first term on the right-hand-side (RHS) is the binary Yukawa repulsion~\cite{KonopkaPRL:2000} between two particles. The second term $-\nu m \boldsymbol{\dot{\mathbf{r}}}_i$ on the RHS represent the frictional gas drag, which is proportional to the velocity of this particle~\cite{B.Liu:2003}. The third term $\zeta_{i}(t)$ on the RHS is the Langevin random kicks coming from the fluctuation-dissipation theorem of $\left\langle\zeta_{i}(0)\zeta_{i}(t)\right\rangle=2m\nu k_{\rm B}T\delta(t)$~\cite{WFVANGunsterenMOLPHYS:1982, FengPRE:2008}. The latter two terms of $\mathbf{F_{\rm s}}$ and $\mathbf{F_{\rm d}}$ represent the forces from the modulation substrate and the external driving force, respectively, both in units of $F_{\rm 0} = Q^2/4 \pi \epsilon_{\rm 0} a^2$, as explained in detail next. 

In our current investigation, following~\cite{youfengweiPRE:2022}, we specify a 1D vibrational periodic substrate as
\begin{equation}\label{EQsubstrate}
      U(x) = \frac{U_{\rm 0}}{(2\pi)^2}\left\{1- \cos\left[\frac{2\pi}{w}\left(x-A\cos \left(2\pi f_{\rm s}t\right)\right)\right]\right\}.
\end{equation}
Here, $U_{\rm 0}$ and $w$ are the strength and width of the applied potential well, in units of $Q^2 / 4 \pi \epsilon_{\rm 0} a$ and $b$, respectively. Different from the previously applied substrates~\cite{LIWPRE:2019, L.GuPRE:2020, zhuwenqiPRE:2022, huangyuPRE1:2022, huangyuPRE2:2022} in dusty plasma investigations, here we apply a lateral periodic excitation $A\cos(2 \pi f_{\rm s} t)$ to the substrate potential~\cite{youfengweiPRE:2022}, where $A$ and $f_{\rm s}$ are the amplitude and frequency of the vibrational excitation. As a result, the substrate force $\mathbf{F_{\rm s}}$ is derived analytically using $-\nabla U(x)$ and Eq.~\eqref{EQsubstrate} as 
\begin{equation}\label{EQsubstrate_force}
	\mathbf{F_{\rm s}} = - \frac{U_{\rm 0}}{2 \pi w}\sin\left[\frac{2 \pi}{w}\left(x-A\cos \left(2 \pi f_{\rm s} t\right)\right)\right]\mathbf{\hat{x}}.
\end{equation}
In our current investigation, we choose the constant values of $U_{\rm 0} = 1.0 Q^2 / 4 \pi \epsilon_{\rm 0} a$, $w = b$, $A = 1.0a$, and $f_{\rm s} = 0.2 \omega_{\rm pd}$ in the substrate expression of Eq.~(\ref{EQsubstrate}), where the nominal dusty plasma $\omega _{\rm pd} = (Q^2/ 2 \pi \epsilon_{\rm 0} m a^{3} )^{1/2}$~\cite{GJKalmanPRL:2004}. The driving force $\mathbf{F_{\rm d}} = F_{\rm d} \mathbf{\hat{x}}$ along the $x$ direction is always specified to be uniform for all simulated particles as in~\cite{LIWPRE:2019, L.GuPRE:2020, zhuwenqiPRE:2022, huangyuPRE1:2022, huangyuPRE2:2022}, whose magnitude varies from 0 to 0.1$F_{\rm 0}$ with the interval of $10^{-3}$$F_{\rm 0}$. 

Other simulation details are listed as follows. We specify the conditions of 2D dusty plasmas as $\Gamma =1000$, $\kappa = 2$, corresponding to the typical solid state~\cite{PHartmannPRE:2005}. Our simulation box is 60.9$a$ $\times$ 52.8$a$ with the periodic boundary conditions, containing $N_{\rm p}=1024$ particles. We guarantee that the periodic boundary conditions are well satisfied for our applied substrate, since the width of our simulation box just corresponds to 32 full potential wells. We specify the frictional gas damping $\nu$ = 0.027$\omega_{\rm pd}$, comparable to the typical value in 2D dusty plasma experiments~\cite{YFengPRE:2011}. We integrate Eq.~\eqref{EQyundong} with a time step of 0.001$\omega_{\rm pd}^{-1}$, small enough for our studied conditions. For each simulation run, after the simulation system reaches its steady state, we record the particles' positions and velocities in the time duration of $\omega_{\rm pd} t = 1.4 \times 10^5$, long enough to obtain the structure and dynamics information. Note, we also perform a few test runs with 4096 particles to confirm that our reported results here are independent of the system size.

\section{Results and discussion}\label{sec3}

\subsection{Shapiro steps}\label{3A}

\begin{figure}
	\centering
	\includegraphics{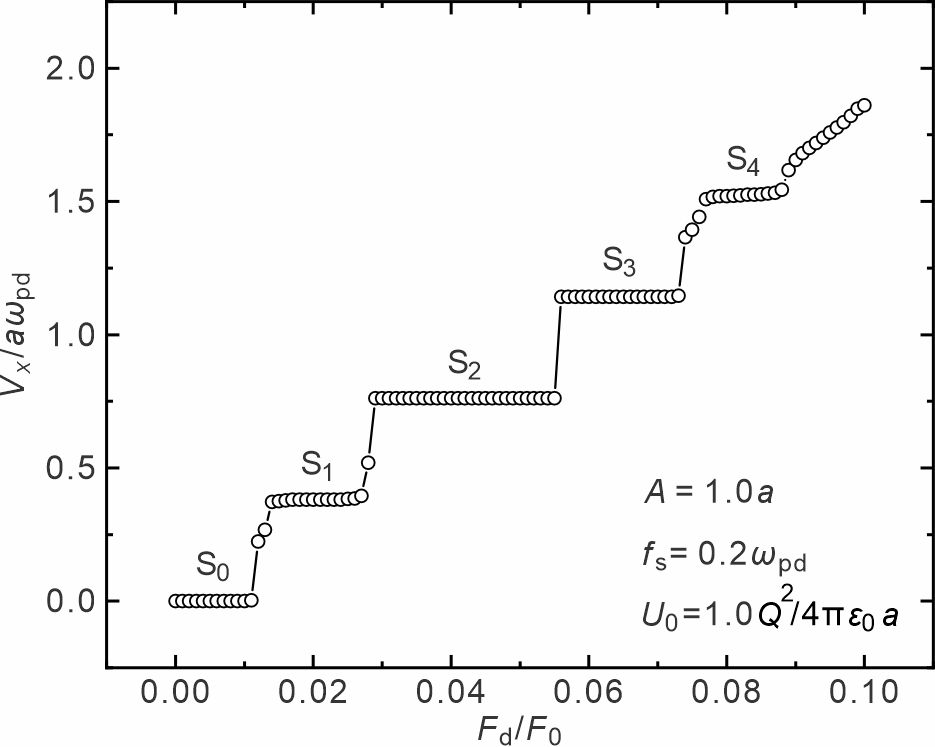}
	\caption{\label{fig1} Calculated collective drift velocity ${V_{x}}$ of a 2D Yukawa solid modulated by a 1D vibrational substrate as a function of a uniform driving force $F_{\rm d}$. As $F_{\rm d}$ increases from zero, the ${V_x}$ results exhibit five distinctive steps, corresponding to ${V_{x}/a\omega_{\rm pd}} = 0$, $0.38$, $0.76$, $1.14$, and $1.52$, labeled as $\rm {\rm S}_{0}$, $\rm {\rm S}_{1}$, $\rm {\rm S}_{2}$, $\rm {\rm S}_{3}$, and $\rm {\rm S}_{4}$, respectively. The first step $\rm {\rm S}_{0}$ corresponds to ${V_{x} / a\omega_{\rm pd}} = 0$, i.e., all particles are confined inside the substrate, indicating the pinned state. While other four steps corresponding to the non-zero ${V_{x}}$ values indicate the typical Shapiro steps, caused by the dynamic mode locking~\cite{TekicPRE:2010, JLAbbottNJphys:2019}. Note, the conditions of the applied 1D vibrational substrate of Eq.~\eqref{EQsubstrate} are $A = 1.0 a$, $f_{\rm s} = 0.2 \omega_{\rm pd}$, and $U_{\rm 0}$ = 1.0 $Q^2 / 4\pi \epsilon_0 a$.
    }
\end{figure}

To investigate the Shapiro step problem in a substrate-modulated 2D Yukawa solid driven by uniform external forces, we first calculate the overall drift velocity along the $x$ direction using $V_{x} = N_{\rm p}^{-1} \left\langle \sum_{i=1}^{N_{\rm p}} {\mathbf{v}_{i} \cdot \hat{\mathbf{x}}} \right\rangle$. Here, $\left\langle\right\rangle$ means the ensemble average, which is achieved using the mean of $2 \times 10^4$ frames for each reported data point. Note, we confirm that the overall drift velocity along the $y$ direction $V_{y}$ is negligible since the applied uniform external force $F_{\rm d}$ is always in the $x$ direction.

As the major result of this paper, from our calculated drift velocity $V_{x}$, we discover four significant Shapiro steps while gradually increasing the uniform driving force on the substrate-modulated 2D Yukawa solid, as presented in Fig.~\ref{fig1}. From Fig.~\ref{fig1}, as $F_{\rm d}$ increases gradually from zero, our $V_{x}$ results exhibit five distinctive steps, corresponding to ${V_{x}/a\omega_{\rm pd}} = 0$, $0.38$, $0.76$, $1.14$, and $1.52$, labeled as $\rm {\rm S}_{0}$, $\rm {\rm S}_{1}$, $\rm {\rm S}_{2}$, $\rm {\rm S}_{3}$, and $\rm {\rm S}_{4}$, respectively. The first step $\rm {\rm S}_{0}$ represents the typical pinned state~\cite{LIWPRE:2019} from $V_{x} = 0$, clearly indicating that all particles are pinned inside potential wells~\cite{LIWPRE:2019}. For a very large driving force of $F_{\rm d}/F_{0} \geq 0.09$, we find that $V_{x}$ increases almost linearly with $F_{\rm d}$ with a fixed slope of $1/{m \nu}$, just corresponding to the typical moving ordered state~\cite{L.GuPRE:2020}, i.e., all particles move as a rigid body over the substrate with a constant speed~\cite{LIWPRE:2019}. Between these Shapiro steps, we also find some data points corresponding to the plastic flow state~\cite{LIWPRE:2019}, as we will study later.

For the four Shapiro steps of $\rm {\rm S}_{1}$, $\rm {\rm S}_{2}$, $\rm {\rm S}_{3}$, and $\rm {\rm S}_{4}$ in Fig.~\ref{fig1}, although the applied driving force $F_{\rm d}$ increases substantially, i.e., $ \geq 0.01 F_{0}$, the overall drift velocity $V_{x}$ remain constant, matching the definition of Shapiro steps~\cite{MPNJunipernc:2015}. From Fig.~\ref{fig1}, the most prominent step is $\rm {\rm S}_{2}$, with the varying range of $0.026 F_{0}$ for $F_{\rm d}$, much wider than those for other three. After $\rm {\rm S}_{2}$, $\rm {\rm S}_{3}$ is still more significant than $\rm {\rm S}_{1}$ and $\rm {\rm S}_{4}$. Note, we also find that, at the two termini of $\rm {\rm S}_{2}$ and $\rm {\rm S}_{3}$, the overall drift velocity $V_{x}$ seems to change abruptly, however, at the two termini of $\rm {\rm S}_{1}$ or $\rm {\rm S}_{4}$, $V_{x}$ varies much more gradually.

Different from the previous studies~\cite{LIWPRE:2019, L.GuPRE:2020, zhuwenqiPRE:2022, huangyuPRE2:2022}, the Shapiro steps discovered in the current investigation are mainly due to the specified lateral periodic excitation $A\cos(2 \pi f_{\rm s} t)$ of Eq.~\eqref{EQsubstrate_force} on the substrate. In Refs.~\cite{LIWPRE:2019, L.GuPRE:2020, zhuwenqiPRE:2022, huangyuPRE2:2022} with the static modulation substrate, when uniform driving forces $F_{\rm d}$ are applied to 2D dusty plasmas or Yukawa systems, the three dynamical states of the pinned, disordered plastic flow, and moving ordered are observed. While in our current investigation, the only difference is that the modulation substrate is vibrational from Eq.~\eqref{EQsubstrate}. As a result, when the driving force $F_{\rm d}$ increases gradually, besides the previously observed three states~\cite{LIWPRE:2019, L.GuPRE:2020, zhuwenqiPRE:2022, huangyuPRE2:2022}, we also find one more dynamic state of the Shapiro steps. 

In fact, Shapiro steps has also been observed in other systems, including Josephson junctions~\cite{YuMShukrionvpRB:2013, VPSpRL:2024}, colloids~\cite{MPNJunipernc:2015, JLAbbottNJphys:2019, TBrazdaSoftmatter:2017}, the Frenkel-Kontorova model~\cite{youfengweiPRE:2022, TekicPRE:2010, wangcanglongPRE:2011, JTekicPRE:2019, BHuPRE:2007}, skyrmions~\cite{JCBSouzaPRB:2024, NPVizarimPRB:2020, CReichardtPRB:2015}, and vortex lattices~\cite{CReichhardtphysicac:2000}. Among these studies, some modulation substrates are vibrational~\cite{TBrazdaSoftmatter:2017, youfengweiPRE:2022, JLAbbottNJphys:2019}, very similar to our 1D vibrational periodic substrate. While in other systems, although the modulation substrate is static, the driving force acting on particles is not uniform any more, which has an typical AC form of $A\cos(2 \pi f_{\rm s} t)$~\cite{YuMShukrionvpRB:2013, VPSpRL:2024, TekicPRE:2010, wangcanglongPRE:2011, JTekicPRE:2019, MPNJunipernc:2015, JCBSouzaPRB:2024, NPVizarimPRB:2020, CReichhardtphysicac:2000, BHuPRE:2007, CReichardtPRB:2015}. From our understanding, to generate Shapiro steps, particles must experience an external AC excitation, either from the modulation substrate, or from the driving force, as we will discuss in detail later. 

\subsection{Dynamic mode locking}\label{3B}

\begin{figure}
	\centering
	\includegraphics{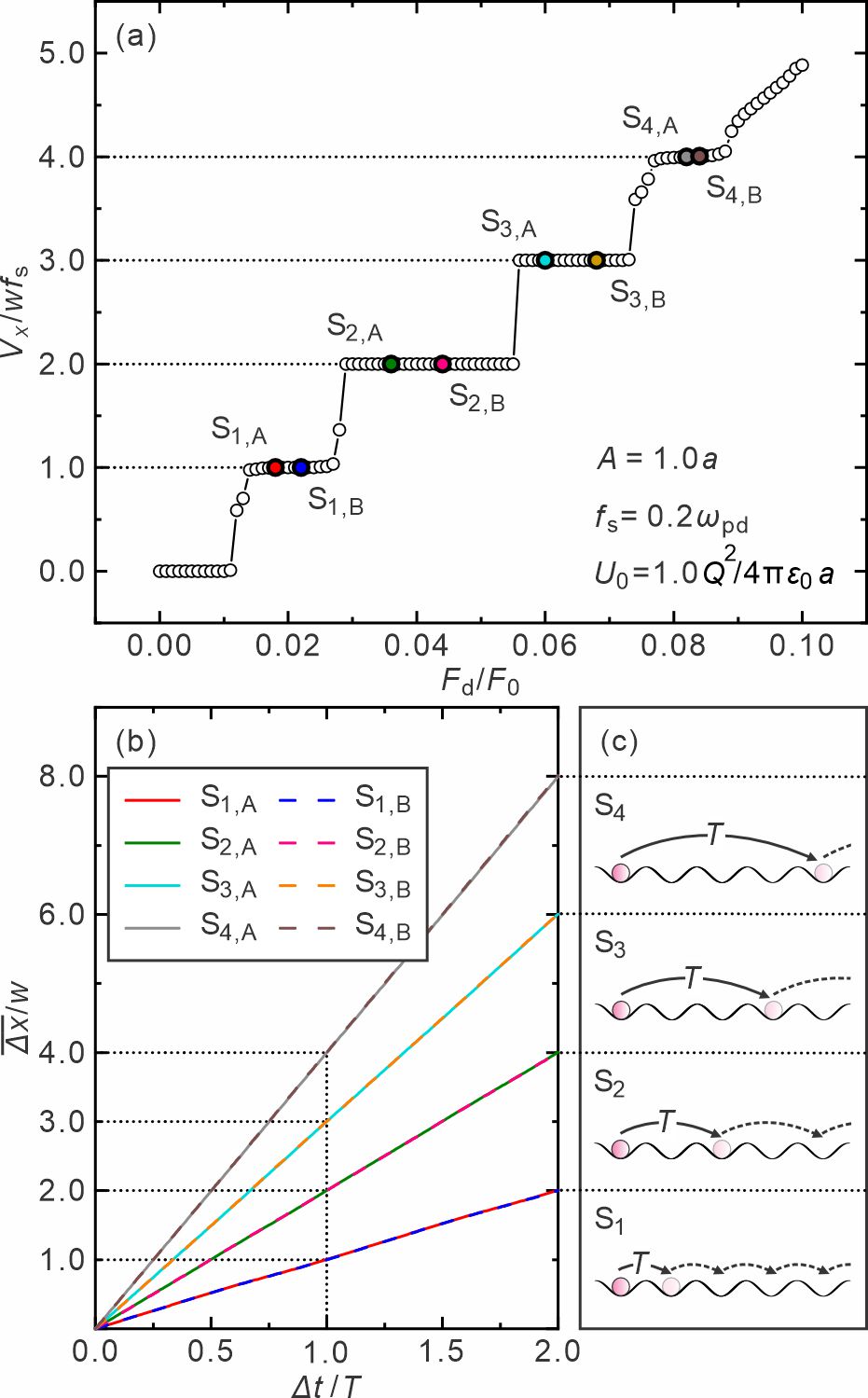}
	\caption{\label{fig2} The $V_{x}$ values from Fig.~\ref{fig1} in units of $wf_{\rm s}$ (a), the corresponding particles' average displacements $\overline{\Delta x}$ (b), and sketches of the microscopic mechanism (c) for different Shapiro steps. Clearly, in units of $wf_{\rm s}$, the Shapiro steps always occur when the $V_{x}$ values are integers, i.e., $V_{x}/(w f_{\rm s}) = 1$, $2$, $3$, and $4$. As a result, the same results of $\overline{\Delta x}$ for ${\rm S}_{n_{\rm s},\rm {A}}$ and ${\rm S}_{n_{\rm s},\rm {B}}$ of each Shapiro step suggest that the average particle displacements are always $n_{\rm s} w$ in one period, as shown in panel (b). Thus, the microscopic mechanism of each observed Shapiro step ${\rm S}_{n_{\rm s}}$ just corresponds to the moving forward $n_{\rm s}$ potential wells in one period for all particles, as illustrated in panel (c).
	} 
\end{figure}

To investigate the underlying mechanism of our discovered Shapiro steps, we replot the calculated $V_x$ results in different units of $wf_{\rm s}$ in Fig.~\ref{fig2}(a). Surprisingly, the discovered four Shapiro steps always occur when $V_x / wf_{\rm s}$ just equals integers, i.e., $V_x =  w f_{\rm s}$, $2  w f_{\rm s}$,  $3  w f_{\rm s}$,  $4  w f_{\rm s}$ for the $1$st, $2$nd, $3$rd, $4$th Shapiro steps, respectively. Note, within each Shapiro step, we always choose different conditions as two marked points ${\rm S}_{n_{\rm s},\rm {A}}$ and ${\rm S}_{ n_{\rm s},\rm {B}}$ for the latter analysis. In Fig.~\ref{fig2}(b), we present the calculated averaged displacement $\overline{\Delta x}$ within two periods under the conditions of our marked points in Fig.~\ref{fig2}(a). From Fig.~\ref{fig2}(b), our $\overline{\Delta x}$ results indicate that the displacement of all particles for the $n_{\rm s}$th Shapiro step in one period is just ${n_{\rm s}}w$, and the motion within each Shapiro step is exactly the same.

To intuitively describe the motion of particles under the conditions of the Shapiro steps, in Fig.~\ref{fig2}(c), we plot the sketch of the microscopic mechanism from our understanding of Fig.~\ref{fig2}(b). For the $n_{\rm s}$th Shapiro step, particles laterally move forward $n_{\rm s}$ potential wells in the first period, then continue moving forward $n_{\rm s}$ potential wells in the second period, and so on. In fact, this microscopic mechanism is believed to be originated from the dynamic mode locking~\cite{TBrazdaSoftmatter:2017}, as discussed next.  

From Ref.~\cite{TKlingerPhysscr:1997}, mode locking, also termed phase locking, refers to the resonance coupling between the free-running oscillator and an external driving force. In fact, dynamic mode locking is also named as synchronization~\cite{TKlingerPhysscr:1997, MPNJunipernc:2015, JodavicPRE:2015, MPNJuniperNEWJphys:2017}. In our current investigation, as the uniform driving force increases, the frequency from the drift motion over potential wells increases simultaneously, while the external frequency $f_{\rm s}$ of the vibrational substrate is specified to be 0.2$\omega_{\rm pd}$. When the ratio of these two frequencies is close to a integer, the results in Fig.~\ref{fig2} clearly indicate the occurrence of dynamic mode locking under the conditions of the four Shapiro steps. Note, due to the synchronized motion of all particles under the Shapiro step conditions, the nominal dusty plasma frequency $\omega_{\rm pd}$ does not play an important role any more. 

\begin{figure}
	\centering
	\includegraphics{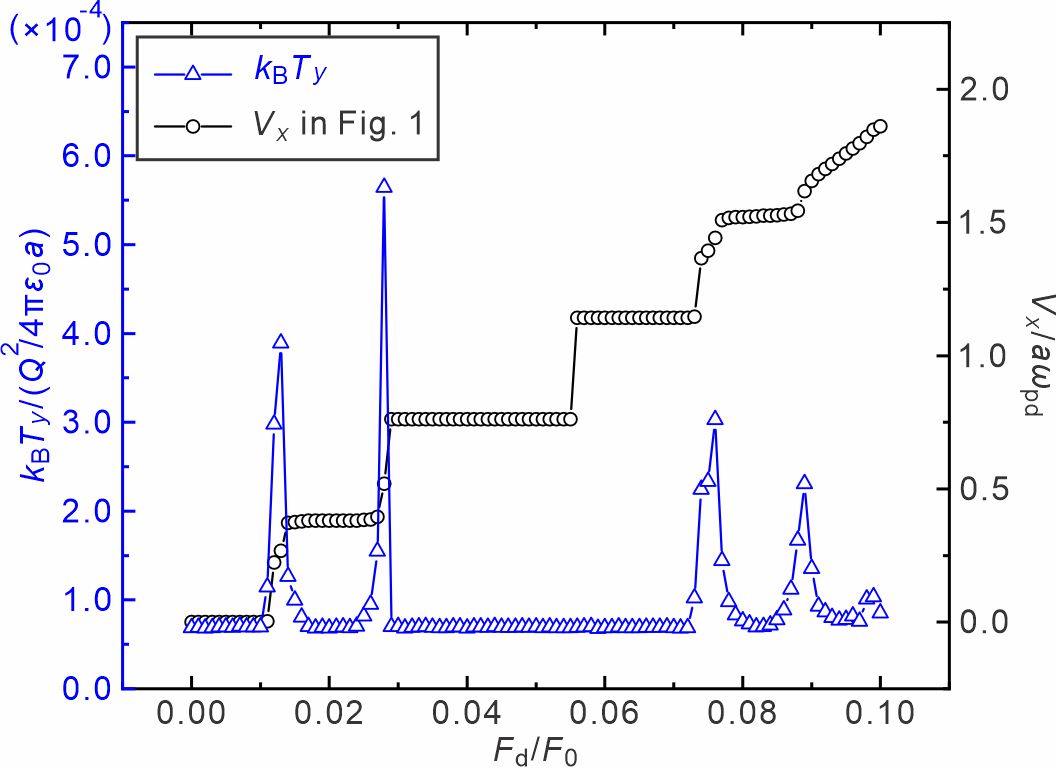}
	\caption{\label{fig3} Calculated kinetic temperature $k_{\rm B}T_{y}$ and the drift velocity $V_{x}$ in Fig.~\ref{fig1} as functions of the driving force $F_{\rm d}$. The $k_{\rm B}T_{y}$ results exhibit four distinctive peaks around both termini of the first and fourth Shapiro steps, i.e., ${\rm S}_{1}$ and ${\rm S}_{4}$. However, under other conditions, the dimensionless kinetic temperature $k_{\rm B}T_{y} / (Q^2 / 4 \pi \epsilon_{\rm 0} a)$ is only around $0.7 \times 10^{-4}$, much lower than values around those four peaks. These four peaks of $k_{\rm B}T_{y}$ probably suggest the continuous transitions related to the first and fourth Shapiro steps.
	} 
\end{figure}

To verify the dynamic mode locking of our observed Shapiro steps, we calculate the kinetic temperature $k_{\rm B}T_{y}$ from the particle motion in the $y$ direction, as presented in Fig.~\ref{fig3}. Here, since our applied modulation substrate is in the $x$ direction and also vibrational in the $x$ direction, the kinetic temperature can be characterized using $k_{\rm B}T_{y} = m\left\langle \sum_{i=0}^{N_{\rm p}}\left(\mathbf{v}_{i,y}-\overline{\mathbf{v}_{y}}\right)^2\right\rangle/2$, in units of $Q^2 / 4 \pi \epsilon_{\rm 0} a$. From Fig.~\ref{fig3}, clearly, there are four distinctive peaks in the profile of $k_{\rm B}T_{y}$, which are just located around both termini of ${\rm S}_{1}$ and ${\rm S}_{4}$. Comparing the kinetic temperature results in Fig.~\ref{fig3} to the drift velocity $V_x$ in Fig.~\ref{fig1}, we find that these four peaks of $k_{\rm B}T_{y}$ just correspond to the disordered plastic flow state at the two termini of ${\rm S}_{1}$ and ${\rm S}_{4}$, respectively. Except for these four peaks, the dimensionless kinetic temperature $k_{\rm B}T_{y} / (Q^2 / 4 \pi \epsilon_{\rm 0} a)$ for other conditions, including the four Shapiro steps, the pinned and moving ordered states, is always pretty low, only around $0.7 \times 10^{-4}$. 

In Fig.~\ref{fig3}, the relatively low level of the $k_{\rm B}T_{y}$ results for the four Shapiro steps and the moving ordered states clearly reflect the velocity fluctuation among particles in the $y$ direction is really tiny, i.e., the motion of all particles are synchronized. In our studied system, since the particle motion is modulated periodically in the $x$ direction by the applied modulation substrate, we cannot use $k_{\rm B}T_{x} = m\left\langle \sum_{i=0}^{N_{\rm p}}\left(\mathbf{v}_{i,x}-\overline{\mathbf{v}_{x}}\right)^2\right\rangle/2$ to determine $k_{\rm B}T_{x}$ directly. However, there is no modulation in the $y$ direction, so that $k_{\rm B}T_{y}$ is able to characterize the kinetic temperature of our studied system. From our understanding, under the conditions of each Shapiro step in Fig.~\ref{fig3}, all particles move along the $x$ direction in a synchronous and periodic fashion~\cite{TBrazdaSoftmatter:2017}, leading to substantially reduced velocity fluctuations, or $k_{\rm B}T_{y}$ equivalently. For the initial pinning and final moving ordered states, $k_{\rm B}T_{y}$ results are almost at the same low level as those for Shapiro steps, since particles either only vibrate around their equilibrium locations or drift completely as a rigid object also in a synchronized manner, as described in detail in Refs.~\cite{LIWPRE:2019, L.GuPRE:2020}.

From our understanding, the four peaks of the $k_{\rm B}T_{y}$ results in Fig.~\ref{fig3} are probably due to the continuous transition around the two termini of ${\rm S}_{1}$ and ${\rm S}_{4}$. From Fig.~\ref{fig3}, under the conditions of the disordered plastic flow state between Shapiro steps, the magnitude of $k_{\rm B}T_{y}$ increases substantially to high values, even nearly $10$ times of the value for the Shapiro steps. In fact, under the conditions of the four peaks of $k_{\rm B}T_{y}$, the results of the drift velocity $V_x$ also exhibit the gradual increase with $F_{\rm d}$, corresponding to the continuous transition there, similar to the results in Ref.~\cite{L.GuPRE:2020}. The only exception is the transition between ${\rm S}_{2}$ and ${\rm S}_{3}$, where $k_{\rm B}T_{y}$ remains at a low level while $V_x$ abruptly changes from ${\rm S}_{2}$ to ${\rm S}_{3}$, probably corresponding to a discontinuous transition~\cite{L.GuPRE:2020}, as discussed later.

\subsection{Continuous and discontinuous transitions}\label{3C}

\begin{figure}
	\centering
	\includegraphics{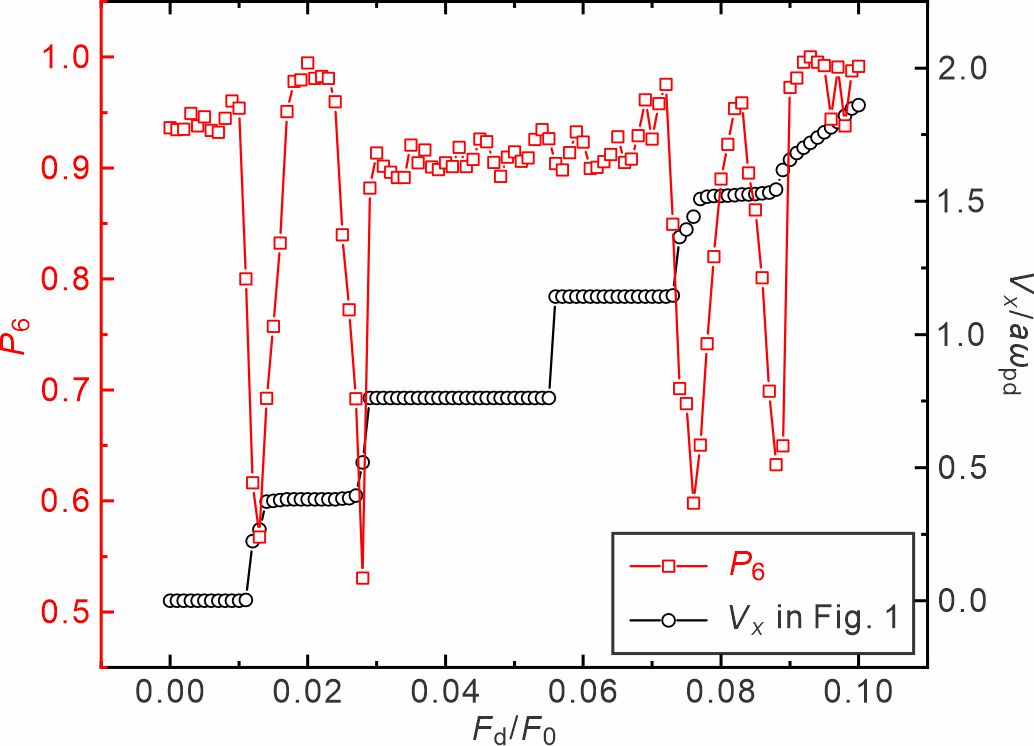}
	\caption{\label{fig4} Calculated results of the fraction of sixfold coordinated particles $P_{6}$ and the drift velocity $V_{x}$ in Fig.~\ref{fig1} as functions of driving force $F_{\rm d}$. The $P_{6}$ results exhibit four significant valleys around both termini of the ${\rm S}_{1}$ and ${\rm S}_{4}$, corresponding to the disordered structure there. From our understanding, these four valleys suggest the continuous transitions related to the first and fourth Shapiro steps. 
	} 
\end{figure}

To investigate the global structure of our studied system, we calculate the fraction of sixfold coordinated particles $P_{6}$. From Ref.~\cite{CReichhardtPRE:2005}, $P_{6}$ is defined as $ P_{6} = N^{-1}_{\rm p}\left\langle \sum_{i=1}^N\delta(6-z_{i}) \right\rangle$, where $z_{i}$ represents the coordination number of particle $i$ determined through the Voronoi construction. When a 2D system varies from a perfect triangular lattice to a highly disordered liquid/gas, its $P_{6}$ value would decay from $1$ to a low value close $0$. We plot our calculated $P_{6}$ and $V_{x}$ results as functions of $F_{\rm d}$ in Fig.~\ref{fig4} for the next analysis.

From Fig.~\ref{fig4}, there are four distinctive valleys in the calculated $P_{6}$ profile, which are just located around both termini of ${\rm S}_{1}$ and ${\rm S}_{4}$. In fact, from the drift velocity $V_x$ profile, these lower values of $P_{6}$ just correspond to the disordering plastic flow state. However, under the conditions of the four Shapiro steps, the pinned and moving ordered states, the $P_{6}$ value is pretty high, mostly higher than $0.9$. 

The relatively high values of $P_{6}$ in Fig.~\ref{fig4} indicate that the structure of our studied 2D Yukawa system is highly ordered. From Fig.~\ref{fig4}, under the conditions of each Shapiro step, the relatively high values of $P_{6}$ clearly indicate the highly ordered structures there. For the conditions of Shapiro steps, the dynamic mode locking of the collective behaviors of particles means the synchronization motion of all particles. From  our understanding, this synchronization reasonably leads to the highly ordered arrangement of particles, as indicated by the high values of $P_{6}$ in Fig.~\ref{fig4}. 

Under other conditions apart from the Shapiro steps, our studied substrate-modulated Yukawa system exhibits the similar depinning dynamics as in~\cite{LIWPRE:2019}. When $F_{\rm d}/F_{0} \leq 0.011$, the $P_{6}$ results are relatively high, while $V_x$ remains nearly zero, corresponding to the pinned state, in which particles are aligned into 1D chains around the bottom of potential wells~\cite{LIWPRE:2019}. When $F_{\rm d}/F_{0} \geq 0.09$, the $P_{6}$ values are also pretty high, $\approx 0.95$, corresponding to the moving ordered state, in which all particles move in an orderly manner just like a rigid object~\cite{LIWPRE:2019}. Interestingly, the four valleys of the $P_{6}$ curves in Fig.4 just correspond to the four transitions from ${\rm S}_{0}$ to ${\rm S}_{1}$, ${\rm S}_{1}$ to ${\rm S}_{2}$, ${\rm S}_{3}$ to ${\rm S}_{4}$, and ${\rm S}_{4}$ to the moving ordered state, which all belong to the plastic flow states~\cite{LIWPRE:2019}. Importantly, for the transition from ${\rm S}_{2}$ to ${\rm S}_{3}$, $P_{6}$ always keeps at high values of $\approx 0.9$. In fact, in our simulations, although we use a much smaller increasing step of $10^{-5}$ for $F_{\rm d}/F_0$, a drop of $P_{6}$ between ${\rm S}_{2}$ to ${\rm S}_{3}$ has never been found, probably indicating the discontinuous transition there. However, other transitions around these Shapiro steps are always continuous instead from the significant drops of $P_{6}$ in Fig.~\ref{fig4}. 

\begin{figure}
	\centering
	\includegraphics{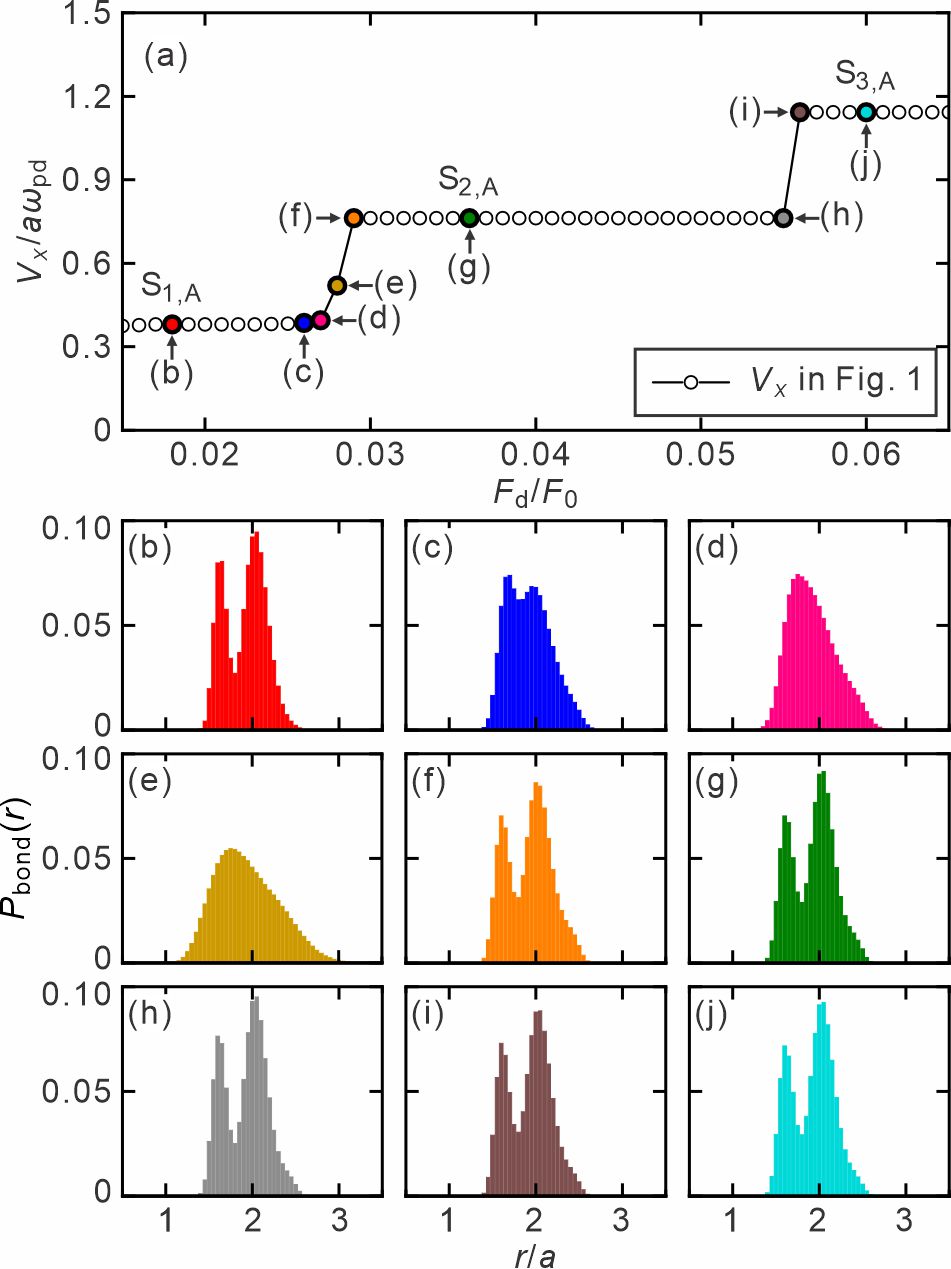}
	\caption{\label{fig5}   Magnified view of the drift velocity $V_{x}$ above (a), and histograms of the bond length $P_{\rm bond} (r)$ (b-j). For the conditions of each Shapiro step, the $P_{\rm bond} (r)$ results exhibit significant bimodal peaks as in panels (b, f, g, h, i, j). However, for the continuous transition conditions, the $P_{\rm bond} (r)$ results exhibit only one prominent peak as in panels (d, e). For the conditions around the first Shapiro step terminus as in panel (c), the $P_{\rm bond} (r)$ results exhibit mainly one peak with a slight feature of bimodal peaks, i.e, the combination of single and bimodal peaks.
	}
\end{figure}

To perform more detailed structure analysis of our studied system around the continuous/discontinuous transitions, we prepare histograms of the bond length $P_{\rm bond} (r)$ using the Delaunay triangulation~\cite{RAQuinnPRE:1996}, as presented in Fig.~\ref{fig5}. In Fig.~\ref{fig5}(a), we mark night points for the conditions of the latter analysis in Figs.~\ref{fig5}(b-j). Clearly, under the conditions on the Shapiro steps as in Figs.~\ref{fig5}(b, f, g, h, i, j), the $P_{\rm bond} (r)$ results exhibit distinctive bimodal peaks. However, under the typical disordered plastic flow state in Figs.~\ref{fig5}(d, e), the $P_{\rm bond} (r)$ results exhibit only one prominent peak. Interestingly, at the right terminus of ${\rm S}_{1}$, the $P_{\rm bond} (r)$ results in Fig.~\ref{fig5}(c) seem to be a transition between one peak and bimodal peaks.

\begin{figure}
	\centering
	\includegraphics{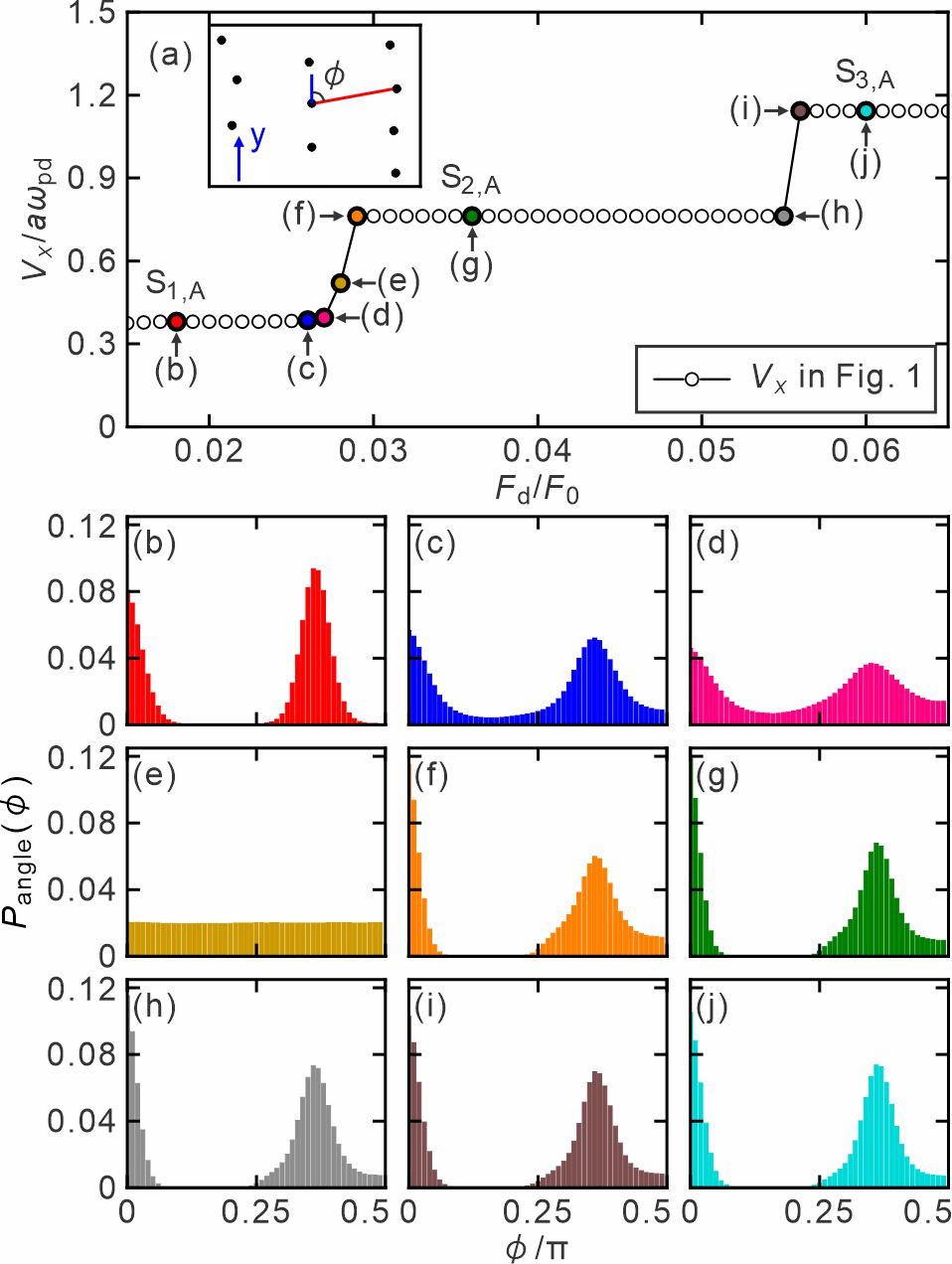}
	\caption{\label{fig6} Magnified view of the drift velocity $V_{x}$ above (a), and histograms of the bond angle $P_{\rm angle} (\phi)$ (b-j). The insert of panel (a) presents a typical snapshot of particle positions (dots) to indicate the definition of the angle $\phi$ of the bond between two neighbors relative to the $y$ axis. For the conditions of each Shapiro step as in panels (b, f, g, h, i, j), the $P_{\rm angle} (\phi)$ results exhibit significant bimodal peaks. For the continuous transition conditions as in panels (c, d), the $P_{\rm angle} (\phi)$ results also exhibit weaker bimodal peaks with the widely distribution of all angles. For the conditions between Shapiro steps as shown in panel (e), $P_{\rm angle} (\phi)$ results exhibit a nearly uniform value for all angles.
    }    
\end{figure}

To further understand the observed bimodal peaks around the Shapiro steps in Fig.~\ref{fig5}, we prepare histograms of the bond angle $P_{\rm angle} (\phi)$ also from the Delaunay 
triangulation~\cite{RAQuinnPRE:1996}, as presented in Fig.~\ref{fig6}. The insert of Fig.~\ref{fig6}(a) shows a local snapshot of particle positions, where the angle between the bond and the $y$ axis is defined as the bond angle $\phi$. Clearly, as in Figs.~\ref{fig6}(b, f, g, h, i, j), the $P_{\rm angle} (\phi)$ results exhibit prominent bimodal peaks, located at $\phi \approx 0$ and $\approx \pi/3$, respectively, while the distribution between $\approx 0.1\pi$ and $\approx 0.25\pi$ is nearly zero. The prominent bimodal peaks in Figs.~\ref{fig6}(b, f, g, h, i, j) clearly indicate the highly ordered structures under these conditions. For the typical disordered plastic flow state of Fig.~\ref{fig6}(e), the bond angle is uniformly distributed, indicating the highly disordered state. Surprisingly, at the right terminus of ${\rm S}_{1}$, the $P_{\rm angle} (\phi)$ result exhibits weaker bimodal peaks, with a wide distribution for all angles, as in Figs.~\ref{fig6}(c, d), corresponding to the ordered structures with some disordered features. 

Our observed bimodal-peak and one-peak features of $P_{\rm bond} (r)$ in Fig.~\ref{fig5} can be well explained by the $P_{\rm angle} (\phi)$ results in Fig.~\ref{fig6}. Under the conditions of Shapiro steps in Figs.~\ref{fig6}(b, f, g, h, i, j), the first peak of $P_{\rm angle} (\phi)$ at $\phi \approx 0$ means that those bonds are nearly in the $y$ directions, corresponding to the arrangement of 1D chains of particles. The second peak at $\phi \approx \pi /3$ means the other bonds are mainly in the direction of $\pm \pi /3$, just corresponding to the stable zigzag structure, formed by neighboring particles from two adjacent 1D chains. We confirm that, under the conditions of the Shapiro steps in Fig.~\ref{fig5}, for the bonds corresponding to the first peak of $P_{\rm bond} (r)$, their angles are mostly $\phi \approx 0$. Similarly, the second peak of the $P_{\rm bond} (r)$ also corresponding to the bonds whose angles are mostly $\phi \approx \pi/3$. As for the typical disordered plastic flow state of Fig.~\ref{fig6}(e), the $P_{\rm angle} (\phi)$ results remain constant, i.e., the bond angles are uniformly distributed, just due to the isotropic property of the highly disordered state. In Figs.~\ref{fig6}(c, d), the $P_{\rm angle} (\phi)$ results exhibit relatively low bimodal peaks, with the widely distribution of bond angles for all values, indicating the combination of the ordered and disordered states, probably leading to the continuous transition there. 

\begin{figure}
	\centering
	\includegraphics{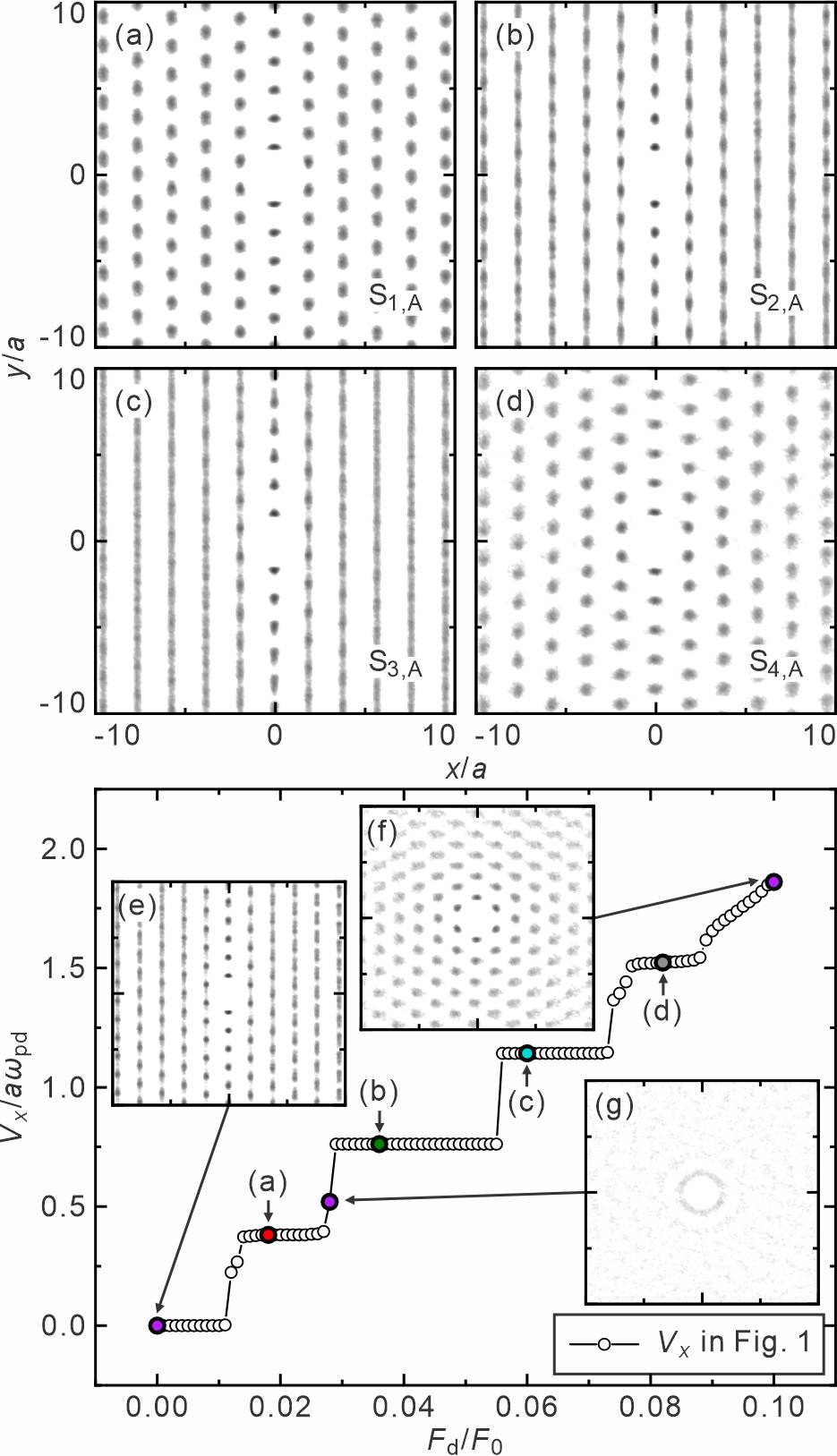}
	\caption{\label{fig7} Calculated 2D distribution functions $G_{xy}$ of our studied 2D Yukawa solid under different conditions of $\rm {\rm S}_{\rm 1,A}$ (a), $\rm {\rm S}_{\rm 2,A}$ (b), $\rm {\rm S}_{\rm 3,A}$ (c), $\rm {\rm S}_{\rm 4,A}$ (d), $F_{\rm d}/F_{0}$ = 0.000 (e), 0.028 (g), and 0.100 (f), respectively. Clearly, the hexagonal symmetry of the $G_{xy}$ results in panels (a, d, f) means that particles are arranged in the highly ordered 2D triangular lattice. While the 1D structure of the $G_{xy}$ results in panels (b, c, e) correspond to the ordered 1D chains. The circular distribution of the $G_{xy}$ results in panel (g) clearly indicates the isotropic uniform distribution of the liquid state, corresponding to the disordered plastic flow state.
    }
    
\end{figure}

To better characterize the 2D arrangement of particles around these Shapiro steps, we calculate 2D distribution functions $G_{xy}$ of our studied 2D system, as presented in Fig.~\ref{fig7}. From Ref.~\cite{KLoudiyiPA:1992}, the $G_{xy}$ function represents the probability of finding a particle at a specific location within a 2D plane relative to a central reference particle, which is powerful to characterize anisotropic systems. The $G_{xy}$ results in Figs.~\ref{fig7}(a, d, f) exhibit the hexagonal symmetry, indicating the highly ordered triangular lattice. The $G_{xy}$ results in Figs.~\ref{fig7}(b, c, e) exhibit the highly ordered structure along the $y$ direction, indicating the 1D chains of the particle arrangement. For the typical disordered plastic flow state, the $G_{xy}$ results exhibit the circular ring-shaped feature, clearly indicating the isotropic property of the highly disordered arrangement of particles, as in Fig.~\ref{fig7}(g). 

From our obtained $G_{xy}$ results in Fig.~\ref{fig7}, we speculate that the two types of the ordered structures of the particle arrangement maybe result in different behaviors observed on the Shapiro steps. Around both termini of ${\rm S}_{1}$ and ${\rm S}_{4}$, the drift velocity
$V_x$ increases gradually, with the corresponding highly ordered triangular lattice with the hexagonal symmetry. However, at the transition between ${\rm S}_{2}$ and ${\rm S}_{3}$, the drift velocity $V_x$ changes abruptly from one step to the other without any intermediate data points at all. As compared with other transitions, the only characteristic is the structures of ${\rm S}_{2}$ and ${\rm S}_{3}$ are both ordered 1D chains. For our studied substrate-modulated Yukawa systems, it seems that any transitions around Shapiro steps either from or to the ordered triangular lattice is always continuous, however, the transition from one ordered 1D-chain structure to the other 1D-chain structure is discontinuous.
              
\section{Summary}\label{sec4}
In summary, we perform Langevin dynamical simulations to study the depinning dynamics of a substrate-modulated 2D Yukawa solid driven by the uniform force. In our investigation, we specify a lateral periodic excitation $A\cos(2 \pi f_{\rm s} t)$ on the applied 1D substrate to introduce an additional frequency into the studied Yukawa system. Besides the previously found pinned, plastic flow, and moving ordered states, we also find that, when the ratio of the frequency from the drift motion over potential wells to the external frequency from modulation substrate is close to integers, dynamic mode locking occurs. As a result, we discover four prominent Shapiro steps from the overall drift velocity of particles. 

We also perform systematic investigations of the structure and dynamics analysis of the 2D Yukawa system around Shapiro steps. From the dynamics analysis, we find that, under the conditions of Shapiro steps, the kinetic temperature $k_{\rm B}T_{y}$ is significantly lower than the plastic flow state, clearly indicating the synchronization motion of all particles. From the global structural measure of sixfold coordinated particles $P_{6}$, we find that, at the discovered Shapiro steps, our studied 2D Yukawa solid is highly ordered. Around both termini of the first and fourth Shapiro steps, the drift velocity increases gradually, corresponds to the continuous transitions. However, between the second and third Shapiro steps, the drift velocity changes from one to the other abruptly, without any data points between, which means the discontinuous transition there. We also prepare the histograms of the length and angle of the bond between neighboring particles under various conditions, which agree with the observed continuous/discontinuous transitions. From our calculated 2D distribution functions, we speculate that the different ordered  arrangements of particles probably result in the continuous or discontinuous transitions between the Shapiro steps and various depinning states.

\subsection*{Acknowledgments}
The work was supported by the National Natural Science Foundation of China under Grant No. 12175159, the 1000 Youth Talents Plan, the Priority Academic Program Development of Jiangsu Higher Education Institutions, and the U. S. Department of Energy through the Los Alamos National Laboratory. Los Alamos National Laboratory is operated by Triad National Security, LLC, for the National Nuclear Security Administration of the U. S. Department of Energy (Contract No. 892333218NCA000001).

\bibliographystyle{apsrev4-1}

\end{document}